\documentclass[11pt,a4paper,notoc]{article}
\usepackage{jheppub}
\usepackage{amsmath}         
\usepackage{amssymb}          
\usepackage{bbold}
\usepackage{ulem}
\usepackage{afterpage}
\usepackage{youngtab}
\usepackage{multirow}
\usepackage[capitalise]{cleveref}
\usepackage{slashed}
\usepackage{booktabs}
\usepackage{subcaption}
\usepackage{comment}
\usepackage{rotating}
\usepackage{url}
\usepackage{physics}
\usepackage{slashbox}
\usepackage{nicematrix}
\usepackage{xcolor}
\usepackage{color}
\usepackage{float}
\definecolor{linkcolor}{rgb}{0,0,0.6}
\definecolor{mygreen}{rgb}{0,0.6,0}
\definecolor{ballblue}{rgb}{0.13, 0.67, 0.8}

\crefname{section}{sec.}{secs.}
\crefname{table}{Tab.}{Tabs.}
\crefname{figure}{Fig.}{Figs.}
\crefname{equation}{Eq.}{Eqs.}
\crefname{appendix}{Appendix}{Appendix}

\newcommand{\SO}{\text{SO}}
\newcommand{\SU}{\text{SU}}
\newcommand{\U}{\text{U}}
\newcommand{\Sp}{\text{Sp}}

\def\beq{\begin{equation}}
\def\eeq{\end{equation}}

\def\gsim{\raise0.3ex\hbox{$\;>$\kern-0.75em\raise-1.1ex\hbox{$\sim\;$}}}
\def\lsim{\raise0.3ex\hbox{$\;<$\kern-0.75em\raise-1.1ex\hbox{$\sim\;$}}}

\usepackage[utf8]{inputenc}

\title{Are there minimal exceptional aGUTs from stable 5D orbifolds?}

\author[a,b]{Giacomo~Cacciapaglia,}
\affiliation[a]{Laboratoire de Physique Th\'eorique et Hautes \'Energies (LPTHE), UMR 7589, Sorbonne Universit\'e \& CNRS, 4 place Jussieu, 75252 Paris Cedex 05, France}
\affiliation[b]{Quantum Theory Center (QTC) \& D-IAS, Southern Denmark Univ., Campusvej 55, 5230 Odense M, Denmark}
\author[c]{Alan~S.~Cornell,}
\affiliation[c]{Department of Physics, University of Johannesburg,
PO Box 524, Auckland Park 2006, South Africa.}
\emailAdd{acornell@uj.ac.za}
\author[d,c]{Aldo~Deandrea,}
\emailAdd{deandrea@ip2i.in2p3.fr}
\affiliation[d]{Universit\'e Claude Bernard Lyon 1, CNRS/IN2P3, IP2I UMR 5822,  4 rue Enrico Fermi, F-69100 Villeurbanne, France}
\author[d]{Wanda~Isnard,}
\emailAdd{wanda.isnard@ens-lyon.fr}
\author[e]{Roman Pasechnik,}
\emailAdd{roman.pasechnik@fysik.lu.se}
\author[e]{Anca~Preda,}
\emailAdd{anca.preda@fysik.lu.se}
\affiliation[e]{Department of Physics, Lund University, SE-223 62 Lund, Sweden}
\author[f]{Zhi-Wei Wang,}
\emailAdd{zhiwei.wang@uestc.edu.cn}
\affiliation[f]{School of Physics, The University of Electronic Science and Technology of China,\\
 88 Tian-run Road, Chengdu, China}

\begin{document}

\abstract{In analysing five dimensional orbifolds with exceptional gauge groups, we seek to find stable vacua configurations which satisfy the minimal requirements for asymptotic grand unified models. In this respect we show that no minimal asymptotic grand unified theory can be built. Our results point towards non-minimal models based on $E_6$: one featuring supersymmetry, and the other needing a modification of the Coleman-Weinberg potential to stabilise the breaking of $E_6$ to the standard model gauge group.}

\maketitle

\section{Introduction}

Although the Standard Model (SM) of particle physics has been tested to high precision in recent years, there are hints pointing towards new physics beyond it. Among these lie the questions regarding the origin of neutrino mass \cite{Athar_2022}, dark matter \cite{Bertone_2018} (and dark energy), and the lightness of the Higgs boson mass \cite{Giudice:2008bi} in spite of seemingly large loop corrections. This prompts the search for new theories that could potentially account for some, if not all, of these puzzles. One elegant and minimal idea is to assume that the SM is superseded by a single, simple gauge group at higher energies, which encompasses the entire SM gauge structure and results in the unification of gauge forces \cite{Georgi}. The conventional paradigm behind the construction of Grand Unified Theories (GUTs) stems from the observation that the gauge couplings converge to similar values at high energies under their renormalisation group evolution. Consequently, quantitative unification is anticipated at a specific scale, $\Lambda_{\rm GUT}$, where the three couplings would meet and beyond which an extended gauge symmetry is restored. The classical and minimal examples are based on $\SU(5)$ \cite{Georgi} or $\SO(10)$ \cite{Georgiso10, FRITZSCH}. At the scale $\Lambda_{\rm GUT}$, the GUT gauge symmetry must be broken through a mechanism analogous to the Higgs mechanism in the SM, typically necessitating numerous scalar fields in large representations. This path of using Higgs-like mechanisms turns out to be a dangerous one for the validity of the theory at high energy, as large representations modify the running of the unified gauge coupling, implying the possible presence of Landau poles not far above the unification scale (especially in supersymmetric GUTs).

Asymptotic unification \cite{Bajc_2016} represents an alternative to this standard picture of unification, from which they mainly differ in the presence of Ultra-Violet (UV) fixed points for the renormalisation evolution of the couplings. Henceforth, in the context of asymptotic Grand Unification Theories (aGUTs) \cite{Cacciapaglia:2020qky}, the three gauge couplings do not cross each other at some high scale, but instead they flow asymptotically towards the same UV fixed point. One way to achieve asymptotic unification is to formulate theories in extra dimensions \cite{Dienes:2002bg}. Perturbative computations allow us to show the emergence of a UV fixed point in $4+\epsilon$ dimensions, which can be extended non-perturbatively up to 5 dimensions (5D) \cite{Gies:2003ic,Morris:2004mg}. In addition, the requirement for an UV fixed point makes these models renormalisable and valid up to high energy scales in the deep UV. Hence, in this work we will focus on 5D theories, where the first examples of aGUT were proposed in Refs.~\cite{Cacciapaglia:2020qky,Cacciapaglia:2023ghp,Cacciapaglia:2023kyz}.

The breaking of gauge symmetry in the context of higher-dimensional gauge theories can proceed via the technique of orbifolding \cite{Kawamura_2001,Hall_2001,Hebecker_2001}. This involves imposing discrete symmetries on the space of the additional space-like dimensions, hence leading to specific boundary conditions for various components of the gauge multiplets. An orbifold is then simply a manifold with singularities where these symmetries act non-trivially \cite{Dixon:1985jw,Kawamura:1999nj, Kachru:2003aw,Kobayashi:2004ya}. The use of orbifolds, therefore, plays a crucial role in reducing large gauge symmetries down to those observed in the SM, and in obtaining massless chiral fermions in the reduced effective 4D theory. 

Another appealing idea of using extra dimensions consists in the gauge-Higgs unification mechanism \cite{HATANAKA_1998}, discussed for both the 5D \cite{von_Gersdorff_2002,Haba_20035D,Haba_20045D} and 6D \cite{Antoniadis_2001, Scrucca_2004, Csaki_2003} cases. It is well known that gauge symmetries can be broken by the presence of extra-dimensional gauge fields, a phenomenon that is called the Hosotani mechanism or Wilson-line symmetry breaking \cite{Hosotani:1983xw,Hosotani:1988bm}.  It relies on the idea that massless zero modes of the extra-space component of the gauge fields, which we call the gauge-scalar, can be identified with the Higgs field \cite{KUBO_2002}. The potential of the gauge-scalar will be protected by the residual gauge invariance after orbifolding and thus, it will be generated radiatively at one loop. Gauge invariance will also ensure the insensitivity of the Higgs mass with respect to the high-scale cut-off, therefore potentially providing an explanation for the lightness of the Higgs boson \cite{Cacciapaglia_2006}.

The construction of realistic aGUT models is not straightforward, as many requirements must be met. A general recipe has been established in Refs.~\cite{Cacciapaglia:2023kyz,Cacciapaglia:2024duu}. In 5D, the most general orbifold $S^1/\mathbb{Z}_2\times\mathbb{Z}_2'$ is defined in terms of two $\mathbb{Z}_2$ parities, $\mathcal{P}_1\times\mathcal{P}_2$, each breaking the bulk gauge group $\mathcal{G}$ to subgroups $\mathcal{H}_1$ and $\mathcal{H}_2$. Hence, the unbroken gauge group in the 4D effective theory is $\mathcal{H} = \mathcal{H}_1 \cap \mathcal{H}_2$. The {\it rules of the game} for the construction of a minimal aGUT is the following: 1) Find the parity such that $\mathcal{H} = \mathcal{G}_{\rm SM} \times \U(1)^n$, where $\mathcal{G}_{\rm SM} = \SU(3)\times\SU(2)_L\times\U(1)_Y$ is the SM gauge group and we allow for additional $\U(1)$ factors; 2) Ensure that the fermion zero modes only contain SM fermions being chiral under $\mathcal{G}_{\rm SM}$; 3) Ensure the existence of UV fixed points for both gauge and Yukawa couplings; 4) Ensure that the orbifold is stable. The last condition relies on the loop-generated potential for the gauge-scalars \cite{Cacciapaglia:2024duu}. A non-minimal extension could allow for an extension of the unbroken 4D gauge group beyond the SM, like, for instance, the Pati-Salam (PS) partial unification \cite{Pati:1974yy}, based on $\SU(4)\times\SU(2)_L\times\SU(2)_R$, which occurs, for instance, in the $\SO(10)$ aGUT \cite{Khojali:2022gcq}. A complete classification for the groups $\SU(N)$, $\Sp(2N)$ and $\SO(N)$ has been presented in Refs.~\cite{Cacciapaglia:2023kyz,Cacciapaglia:2024duu}, showing that only three models pass all requirements: one based on $\SU(6)$, and two PS ones based on $\SO(10)$ and $\SU(8)$.  In this work, we will complete the classification by analysing exceptional groups, $G_2$, $F_4$, $E_6$, $E_7$ and $E_8$. It should be noted that $G_2$ and $F_4$ do not accommodate the SM gauge symmetry \cite{Slansky:1981yr}, while $E_8$ can only be broken to real subgroups via $\mathbb{Z}_2$ parities \cite{Hebecker:2003jt}, hence only $E_6$ and $E_7$ offer interesting aGUT possibilities. One supersymmetric aGUT model based on $E_6$ has been presented in Ref.~\cite{Cacciapaglia:2023ghp}.

The paper is organised as follows: In Section~\ref{sec:crit} we define our strategy to find stable orbifolds for the exceptional groups, based on finding the maximal common subgroups shared by the two parities. In Section~\ref{sec:exceptional} we apply the analysis to all exceptional groups, and we focus on two examples based on $E_6$ and $E_7$, which have some interest for aGUT model building. Finally, we present our conclusions and perspectives in Section~\ref{sec:concl}. Other cases and more details are relegated to the appendices.

\section{The maximal subgroup criterion} \label{sec:crit}

We consider the most general 5D orbifold, $S^1/\mathbb{Z}_2\times\mathbb{Z}_2'$, defined in terms of two independent $\mathbb{Z}_2$ parities. Each parity breaks the bulk gauge group $\mathcal{G}$ to a regular maximal subgroup $\mathcal{H}$. 
As discussed in Ref.~\cite{Hebecker:2003jt}, the parities and symmetry-breaking patterns can be classified via the subgroups. As shown there, every maximal regular subgroup can be generated by an orbifold twist. This allows us to label all possible parities $\mathcal{P}_i$ in terms of their unbroken maximal subgroups $\mathcal{H}_i \subset \mathcal{G}$. We indicate with $\mathcal{P}_0$ the identity, for which $\mathcal{H}_0 = \mathcal{G}$.
For the most general orbifold $S^1/\mathbb{Z}_2\times\mathbb{Z}_2'$, one needs to find the unbroken group $\mathcal{H} = \mathcal{H}_i \cap \mathcal{H}_j$ that stems from a stable orbifold, i.e. it is not destabilised by a vacuum expectation value (VEV) of the massless gauge-scalars. For the ordinary groups $\SU(N)$, $\Sp(2N)$ and $\SO(N)$, a complete classification has been provided in Ref.~\cite{Cacciapaglia:2024duu} by studying all the possible alignments of the parity projection matrices within the group $\mathcal{G}$. This procedure, however, can be rather involved for the exceptional groups. Hence, here we propose to only study cases where $\mathcal{H}$ is a maximal subgroup of both $\mathcal{H}_i$ and $\mathcal{H}_j$. As we will prove here, this procedure allows us to find all the stable orbifolds for the non-exceptional groups listed in Ref.~\cite{Cacciapaglia:2024duu}. 

In all generality, we can identify three template situations:
\begin{itemize}
    \item[A)] For $\mathcal{P}_0 \times \mathcal{P}_i$, the maximal subgroup is $\mathcal{H}_i$, without gauge-scalars. No other alignment is possible; hence orbifolds of this type are always stable.
    \item[B)] For $\mathcal{P}_i \times \mathcal{P}_i$ with $i\neq 0$, the maximal subgroup is $\mathcal{H}_i$ with gauge-scalars in the coset $\mathcal{G}/\mathcal{H}_i$, consisting of all components of the adjoint of $\mathcal{G}$ that are not in the adjoint of $\mathcal{H}_i$. It is easy to check that all components of the adjoint of $\mathcal{G}$ have parities $(+,+)$ or $(-,-)$, hence their contribution to the gauge-scalar potential has a minimum at zero \cite{Cacciapaglia_2006,Cacciapaglia:2024duu}. All orbifolds of this type are, therefore, always stable.
    \item[C)] For $\mathcal{P}_i \times \mathcal{P}_j$ with $i\neq j$, the maximal common subgroup may not be uniquely identifiable. Hence, the stable orbifold alignment of the two parities needs to be studied case by case.
\end{itemize}
We checked, by considering only the maximal common subgroups, that it is possible to identify all the stable orbifolds for the groups $\SU(N)$, $\Sp(2N)$ and $\SO(N)$. We provide below some details for $\SU(N)$, where the other groups offer a trivial generalisation. In the next section, we will apply the same strategy to the exceptional groups.

\subsection{The $\SU(N)$ example}

For $\SU(N)$, there exist $[N/2]$~\footnote{$[N/2]$, the integer part of $N/2$, is equal to $n$ for $\SU(2n)$ and $\SU(2n+1)$ groups.} non-trivial parities $\mathcal{P}_A$ with $A = N-[N/2], \dots, N-1$. The action of the parity on the fields is characterised by a diagonal parity matrix with $A$ entries $+1$ and $N-A$ entries $-1$. Each $\mathcal{P}_A$ parity breaks the $\SU(N)$ gauge group on a boundary:
\begin{equation}
    \SU(N) \to \frac{\U(A) \times \U(N-A)}{\U(1)}\,,
\end{equation}
with
\begin{equation}
    \text{Adj} \to (\text{Adj}, 1) \oplus (1,\text{Adj}) \oplus (F, \bar{F}) \oplus (\bar{F}, F)\,, 
\end{equation}
where $F$ and Adj indicate the fundamental and the adjoint representation of the corresponding group.
Note that we have defined $A\geq N/2$ to avoid double counting, as the overall sign of the parity matrix is not important.

It was shown in Ref.~\cite{Cacciapaglia:2024duu} that the only stable orbifolds feature either two or three $\SU$ factors, with specific gauge-scalar spectra. For the three-block case, stable orbifolds always feature one $\SU(X)$ block with $X \geq N/2$ and a gauge-scalar transforming in its fundamental representation.  The two-block cases correspond trivially to the orbifold types A) and B), which are always stable. Three-block orbifold, instead, can be obtained as in C) by combining two different parities.
When combining $\mathcal{P}_A \times \mathcal{P}_B$ with $A\neq B$, the unbroken subgroup can be identified by the alignment of the $\pm 1$ entries in the two parity matrices. The maximal common subgroup is obtained by aligning the signs in the largest blocks of the two, of dimension $A$ and $B$ respectively. Hence, assuming for simplicity $A>B$, we obtain:
\begin{equation}
    \mathcal{H} = \frac{\U(B) \times \U(A-B) \times \U(N-A)}{\U(1)}\,,
\end{equation}
with a gauge-scalar $\varphi = (F,1,\bar{F})$.
This configuration exactly matches the stable three-block case in Ref.~\cite{Cacciapaglia:2024duu}, as $B \geq N/2$. Hence, all stable $\SU(N)$ orbifolds match the criterion of unbroken maximal subgroup.

\section{Stable exceptional orbifolds} \label{sec:exceptional}

\begin{table}[h!] \centering
\begin{tabular}{|l|c|c|c|}
\hline
Group & Parity & Unbroken group & Adjoint decomposition \\ \hline \hline
$G_2$ & $\mathcal{P}_1$ & $\SU(2)\times\SU(2)$ & $(3,1) \oplus (1,3) \oplus \textcolor{blue}{(2,4)}$ \\ \hline\hline
$F_4$ & $\mathcal{P}_1$ & $\SO(9)$ & $36 \oplus \textcolor{blue}{16}$ \\ \cline{2-4}
& $\mathcal{P}_2$ & $\Sp(6) \times \SU(2)$ & $(21,1) \oplus (1,3) \oplus \textcolor{blue}{(14',2)}$ \\ \hline\hline
$E_6$ & $\mathcal{P}_1$ & $\SO(10)\times\U(1)$ & $45_0 \oplus 1_0 \oplus \textcolor{blue}{16_{-3}} \oplus \textcolor{blue}{\overline{16}_3}$ \\ \cline{2-4}
& $\mathcal{P}_2$ & $\SU(6)\times\SU(2)$ & $(35,1) \oplus (1,3) \oplus \textcolor{blue}{(20,2)}$ \\\hline\hline
$E_7$ & $\mathcal{P}_1$ & $\SU(8)$ & $63 \oplus \textcolor{blue}{70}$ \\ \cline{2-4}
& $\mathcal{P}_2$ & $\SO(12)\times\SU(2)$ & $(66,1) \oplus (1,3) \oplus \textcolor{blue}{(32^{(\prime)},2)}$ \\ \cline{2-4}
& $\mathcal{P}_3$ & $E_6\times\U(1)$ & $78_0 \oplus 1_0 \oplus \textcolor{blue}{27_4} \oplus \textcolor{blue}{\overline{27}_{-4}}$ \\ \hline\hline
$E_8$ & $\mathcal{P}_1$ & $\SO(16)$ & $120 \oplus \textcolor{blue}{128}$ \\ \cline{2-4}
& $\mathcal{P}_2$ & $E_7 \times \SU(2)$ & $(133,1) \oplus (1,3) \oplus \textcolor{blue}{(56,2)}$ \\  \hline
\end{tabular}
\caption{\label{tab:Plist} List of all breaking patterns for the exceptional groups generated by a $\mathbb{Z}_2$ parity \cite{Hebecker:2003jt}. Note that for $E_7$ broken by $\mathcal{P}_2$, the $32$ can be either the spinorial or the $32'$ representation, indicating two possible alignments. The representations in the coset are highlighted in \textcolor{blue}{blue}.}
\end{table}

All $\mathbb{Z}_2$ parities and the related symmetry-breaking patterns of exceptional groups have been listed in Ref.~\cite{Hebecker:2003jt}, which is used to construct Table~\ref{tab:Plist}, including the decomposition of the adjoint representation \cite{Slansky:1981yr}. In Table~\ref{tab:Olist} we compile all the maximal unbroken subgroups obtainable by combining two parities. We see that for the rank-two group, $G_2$, only one parity is defined, hence all maximal orbifolds are stable. For the rank-four group, $F_4$, two parities are defined. The only nontrivial combination has a single maximal subgroup, $\Sp(4)\times\SU(2)\times \SU(2)$. We explicitly computed the potential for the gauge-scalar in the $(4,1,2)$ representation to check the stability of such configuration, see Appendix~\ref{app:F4}. 

\begin{table}[h!]\centering
\begin{tabular}{|l|c|c|c|c|}
\hline
Group & Parities & Unbroken group & Gauge-scalar & Stable? \\ \hline \hline
$G_2$ & $\mathcal{P}_{0/1} \times \mathcal{P}_1$ & $\SU(2) \times \SU(2)$ & None/$(2,4)$ & Yes\\  \hline\hline
$F_4$ & $\mathcal{P}_{0/1} \times \mathcal{P}_1$ & $\SO(9)$ & None/$16$ & Yes \\ \cline{2-5}
& $\mathcal{P}_{0/2} \times \mathcal{P}_2$ & $\Sp(6)\times \SU(2)$ & None/$(14',2)$ & Yes \\ \cline{2-5}
& $\mathcal{P}_{1} \times \mathcal{P}_2$ & $\Sp(4)\times \SU(2)\times \SU(2)$ & $(4,1,2)$ & Yes \\ \hline\hline 
$E_6$ & $\mathcal{P}_{0/1}\times \mathcal{P}_1$ & $\SO(10) \times \U(1)$ & None/$16_{-3}$ & Yes \\ \cline{2-5}
& $\mathcal{P}_{0/2}\times \mathcal{P}_2$ & $\SU(6) \times \SU(2)$ & None/$(20,2)$ & Yes \\\cline{2-5}
& $\mathcal{P}_{1}\times \mathcal{P}_2$ & $\SU(5) \times \U(1) \times \U(1)$ & $10_{-1,-3}$ & Yes \\\cline{3-5}
& & $\SU(4) \times \SU(2) \times \SU(2) \times \U(1)$ & $(4,1,2)_{3}$ & \textcolor{red}{No} \\\hline\hline
$E_7$ & $\mathcal{P}_{0/1}\times \mathcal{P}_1$ & $\SU(8)$ & None/$70$ & Yes \\ \cline{2-5}
& $\mathcal{P}_{0/2}\times \mathcal{P}_2$ & $\SO(12)\times \SU(2)$ & None/$(32,2)$ & Yes \\ \cline{2-5}
& $\mathcal{P}_{0/3}\times \mathcal{P}_3$ & $E_6\times \U(1)$ & None/$27_4$ & Yes \\ \cline{2-5}
& $\mathcal{P}_{1}\times \mathcal{P}_2$ & $\SU(4)\times\SU(4)\times\U(1)$ & $(4,\bar{4})_{-2} $ & Yes \\ \cline{3-5}
&& $\SU(6)\times\SU(2)\times\U(1)$ & $(20,2)_{0} $ & \textcolor{red}{No} \\ \cline{2-5}
& $\mathcal{P}_{1}\times \mathcal{P}_3$ & $\SU(6)\times\SU(2)\times\U(1)$ & $(15,1)_{4} $ & Yes \\ \cline{2-5}
& $\mathcal{P}_{2}\times \mathcal{P}_3$ & $\SU(6)\times\SU(2)\times\U(1)$ & $(6,2)_{-4}$ & \textcolor{red}{No} \\ \cline{3-5}
&& $\SO(10)\times\U(1)\times\U(1)$ & $16_{1,4}$ & Yes\\ \hline\hline
$E_8$ & $\mathcal{P}_{0/1} \times \mathcal{P}_1$ & $\SO(16)$ & None/$128$ & Yes \\ \cline{2-5}
& $\mathcal{P}_{0/2} \times \mathcal{P}_2$ & $E_7 \times \SU(2)$ & None/$(56,2)$ & Yes \\ \cline{2-5}
& $\mathcal{P}_{1} \times \mathcal{P}_2$ & $\SO(12)\times\SU(2)\times\SU(2)$ & $(32,1,2)$ & - \\ \cline{3-5}
&& $\SU(8)\times\U(1)$ & $28_{-1}$ & - \\ \hline
\end{tabular}
\caption{\label{tab:Olist} List of all maximal subgroups obtained by combining two parities. When two possibilities are present, we find that only one is stable (the unstable one is highlighted in \textcolor{red}{red}). The $E_8$ case is included for completeness, though we do not study its stability due to the lack of relevance for aGUT model building.}
\end{table}

For the higher rank groups, $E_6$, $E_7$ and $E_8$, in many cases two possible maximal subgroups can be defined, and we expect only one of these to give a stable parity alignment. Hence, an explicit computation of the gauge-scalar potential is necessary to determine which one corresponds to a stable orbifold. As shown in Table~\ref{tab:Olist}, these cases are $E_6$ with $\mathcal{P}_1\times\mathcal{P}_2$, $E_7$ with $\mathcal{P}_1\times\mathcal{P}_2$ and $\mathcal{P}_2\times\mathcal{P}_3$, and $E_8$ with $\mathcal{P}_1\times\mathcal{P}_2$. We explicitly checked the stability of the $E_6$ and $E_7$ cases, with the answer indicated in the last column of Table~\ref{tab:Olist}. In the following we focus on the two cases that are relevant for aGUT model building, as we will discuss below, i.e. $E_6$ with $\mathcal{P}_1\times\mathcal{P}_2$ and $E_7$ with $\mathcal{P}_2\times\mathcal{P}_3$. We do not consider further the $E_8$ case, as the unbroken groups are always real, and they do not allow for the inclusion of chiral fermion content as present in the SM.

Before introducing the two examples, it is convenient to recap the main notation for the gauge-scalar potential computation. We refer the reader to Ref.~\cite{Cacciapaglia:2024duu} and references therein for more details. The VEVs of the gauge-scalars can always be expressed in terms of `angles' $a_i$, where the potential has a shift symmetry by integers, $a_i \to a_i + k$ with $k\in \mathbb{Z} $. The potential is expressed in terms of the contribution of bulk gauge bosons, fermions, and scalars as follows:
\begin{equation} \label{eq:potgen}
    V_{\rm eff} (a_i) = C \times \left[-3\ \mathcal{V}_{R_G} + 4\ \sum_f \mathcal{V}_{R_f} - \sum_s \mathcal{V}_{R_s} \right]\,,
\end{equation}
where the functions $\mathcal{V}$ only depend on the representations of the gauge adjoint $G$, and the bulk fermions $f$ and real scalars $s$. Here, $C={1}/{(32 \pi^6 R^4)}$ is a positive normalisation constant. The functions $\mathcal{V}$ can be expressed in terms of two functional templates:
\begin{equation}
    F^+ (a) = \frac{3}{2} \sum_{n=1}^\infty \frac{\cos (2\pi n a )}{n^5}
\end{equation}
stemming from components with parities $(\pm,\pm)$, and
\begin{equation}
    F^- (a) = \frac{3}{2} \sum_{n=1}^\infty (-1)^n \frac{\cos (2\pi n a )}{n^5} = - F^+ (a) + \frac{1}{16} F^+(2a)
\end{equation}
stemming from components with parities $(\pm,\mp)$.
It is important to recall that $-F^+(a)$ has a minimum at $a=0$, while $-F^-(a)$ has a minimum at $a=1/2$. Hence, while the former tends to stabilise the orbifold, the latter destabilises it.

\subsection{An $E_6$ example}
\label{subsec:E6}

The breaking of $E_6$ by $\mathcal{P}_1\times\mathcal{P}_2$ is of relevance, as one possible maximal subgroup $\mathcal{H}$ consists of PS symmetry, $\SU(4)\times\SU(2)_L\times\SU(2)_R$, with an additional $\U(1)_\psi$ factor \cite{Gursey:1975ki}, hence it could be used to construct aGUT models.
We recall that the two parities break:
\begin{eqnarray}
    \mathcal{P}_1 & \Rightarrow & E_6 \to \SO(10) \times \U(1)_\psi\,, \\
    \mathcal{P}_2 &\Rightarrow & E_6 \to \SU(6)
    \times \SU(2)_R\,,
    \label{eq:E6toSU6}
\end{eqnarray}
where we have identified the unbroken $\SU(2)$ with $\SU(2)_R$. Since the gauge-scalar transforms as a doublet under the unbroken $\SU(2)$ present in Eq.~\eqref{eq:E6toSU6}, it could be used to reduce the PS symmetry to the SM one, by breaking $\SU(2)_R$.
From Table~\ref{tab:Olist}, we see that there are two possible maximal subgroups that could correspond to the stable parity alignment:
\begin{itemize}
    \item[-] $\SU(5) \times \U(1)_\chi \times \U(1)_\psi$, where $\SU(5) \times \U(1)_\chi \subset \SO(10)$. The fundamental and adjoint of $E_6$ decompose in the following way, where we also indicate the intrinsic parities under $\mathcal{P}_1\times\mathcal{P}_2$:
    \begin{eqnarray}
        {\bf 27} &\to & 10_{-1,1}^{++} \oplus \overline{5}^{+-}_{3,1} \oplus 1^{+-}_{-5,1} \oplus 5^{-+}_{2,-2} \oplus \overline{5}^{--}_{-2,-2} \oplus 1^{--}_{0,4} \,, \\
        {\bf 78} &\to & 24^{++}_{0,0} \oplus 1^{++}_{0,0} \oplus 1^{++}_{0,0} \oplus 10^{+-}_{4,0} \oplus \overline{10}^{+-}_{-4,0} \oplus 10^{--}_{-1,-3} \oplus \overline{10}^{--}_{1,3} \oplus \nonumber \\
        &&  \overline{5}^{-+}_{3,-3} \oplus 5^{-+}_{-3,3} \oplus 1^{-+}_{-5,-3} \oplus 1^{-+}_{3,5}\,.
    \end{eqnarray}
    Hence, this pattern has a gauge-scalar transforming as $\varphi_{\rm SU5} = 10_{-1,-3} + \mbox{c.c.}$.

    \item[-] $\SU(4) \times \SU(2)_L \times \SU(2)_R \times \U(1)_\psi$, where $\SU(4)\times\SU(2)_L \times \U(1)_\psi \subset \SU(6)$. The fundamental and adjoint decompose as
    \begin{eqnarray}
        {\bf 27} &\to & (4,2,1)^{++}_1 \oplus (\overline{4},1,2)^{+-}_1 \oplus (6,1,1)^{-+}_{-2} \oplus (1,2,2)^{--}_{-2} \oplus (1,1,1)^{-+}_{4}\,, \label{eq:27deco}\\
        {\bf 78} &\to & (15,1,1)^{++}_{0} \oplus (1,3,1)^{++}_{0} \oplus (1,1,3)^{++}_{0} \oplus (1,1,1)^{++}_{0} \oplus (6,2,2)^{+-}_{0} \oplus \nonumber \\
        && (4,2,1)^{-+}_{-3} \oplus (\overline{4},2,1)^{-+}_{3} \oplus (4,1,2)^{--}_{3} \oplus (\overline{4},1,2)^{--}_{-3}\,. \label{eq:78deco}
    \end{eqnarray}
    In this case, the gauge-scalar transforms as $\varphi_{\rm PS} = (4,1,2)_3 + \mbox{c.c.}$. This identification of the $\SU(2)$ factors was chosen in such a way that the gauge-scalar vacuum expectation value could be used to break the PS group  to the SM one \cite{Pati:1974yy}.
\end{itemize}

\begin{figure}[H]
    \centering
    \includegraphics[height=6.3cm]{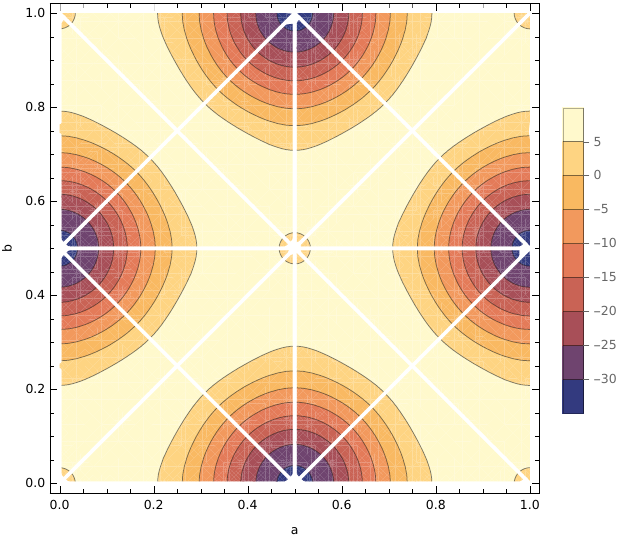} \hfill \includegraphics[height=6.3cm]{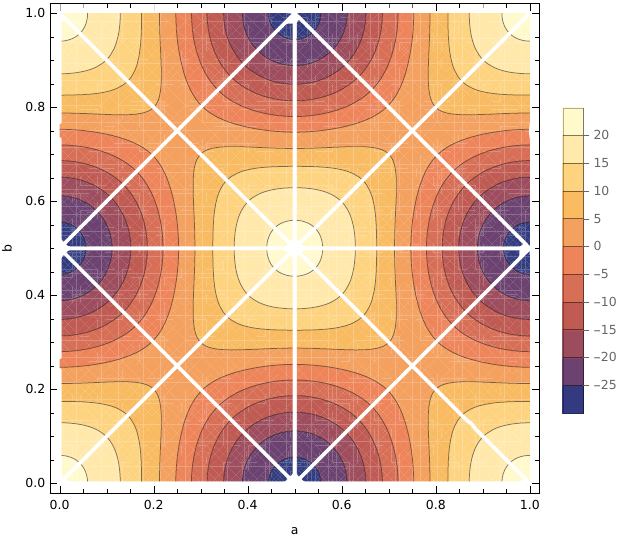}
    \caption{Plot of the gauge contributions to the effective potential (left) and the full effective potential (right) in function of the two VEVs $a$ and $b$.}
    \label{fig:Potential3D}
\end{figure}

To check which configuration yields a stable orbifold, it is necessary to compute the one-loop effective potential for the gauge scalars. We find it most convenient to compute it for the PS configuration; see Appendix~\ref{app:E6} for more details. The gauge scalar $\varphi_{\rm PS}$ allows for two independent vacuum expectation values, which we label $a$ and $b$. The contribution of the adjoint gauge multiplet reads:
\begin{equation}
    \mathcal{V}_{\rm Adj} = \frac{5}{4} \left[ F^+ (2a) + F^+ (2b) \right] + \frac{1}{8} \left[ F^+(2a+2b) + F^+ (2a-2b)\right] + 2 \left[ F^-(a+b) + F^- (a-b)\right]\,.
\end{equation}
Contours of the two-dimensional potential are shown in the left panel of Fig.~\ref{fig:Potential3D}. The two global minima non-related by the periodicity are $(0,1/2)$ and $(1/2,0)$. We also note here the presence of local minima at $(0,0)$ and $(1/2,1/2)$. At the global minima, the PS group is broken to the SM one, however, due to the maximal value of the VEV, additional zero modes appear, which reconstruct the $\SU(5)$ invariance. Hence, the unbroken gauge symmetry at the global minima is $\SU(5)\times \U(1)^2$, hence demonstrating that this is the stable orbifold.

The above conclusion holds based only on the contribution of the gauge fields. However, before completely discarding the PS symmetry breaking, one would need to try and construct a realistic aGUT model including bulk fermion and scalar fields, following the rules established in Ref.~\cite{Cacciapaglia:2023kyz}. We recall that under Pati-Salam, one generation of SM fermions transforms as a left-handed $(4,2,1)$ and a right-handed $(4,1,2)$, while the SM Higgs is embedded in a $(1,2,2)$. From Eq.~\eqref{eq:27deco}, it is clear that the minimal field content consists of fundamentals of $E_6$.\footnote{A non-minimal scenario could be obtained by embedding the left-handed fermions in the adjoint~\cite{Kobayashi:2004ya,Cacciapaglia:2023ghp}.} For matter fields, we can assign overall parities, which flip the intrinsic parities listed in Eq.~\eqref{eq:27deco}: hence, the left-handed fermions can be obtained from a $\Psi_{\bf 27}^{++}$, while the right-handed ones from $\Psi^{-+}_{\bf \overline{27}}$. The Higgs instead emerges from a scalar $\Phi_{\bf 27}^{--}$. One can check that the additional zero modes transform under real representations of the PS group; hence they acquire mass once the $\U(1)_\psi$ is broken.\footnote{Note that the model has 4D anomalies involving the $\U(1)_\psi$ current, which can be cancelled by localised chiral fermions and/or localised Chern-Simons terms \cite{Scrucca:2001eb,Barbieri:2002ic,Scrucca:2004jn}.} The field content allows for one Yukawa coupling in the bulk:
\begin{equation}
    \mathcal{L}_{\rm Yuk} = Y\ \overline{\Psi}_{\bf \overline{27}} \Psi_{\bf 27} \Phi_{\bf 27} + \mbox{h.c.}
\end{equation}
As such, we have computed the renormalisation group equations for the gauge and Yukawa couplings in 5D \cite{Cacciapaglia:2023kyz}, and found that for $n_g$ ``families'' of fermions, the following fixed points exist:
\begin{equation}
    \alpha_g^\ast = \frac{2\pi}{41-8n_g}\,, \qquad \alpha_y^\ast = \frac{(22+16n_g)\pi}{135(41-8n_g)}\,,
\end{equation}
where $\alpha_x = x^2/(4\pi)$. These UV fixed points are well defined as long as $n_g \leq 5$.
Adding the contribution of scalars and three generations of fermions to the gauge-scalar potential, we obtain the right panel of Fig.~\ref{fig:Potential3D}. This confirms the global minima at the $\SU(5)\times\U(1)^2$ orbifold. 

Note, finally, that if the gauge-scalar VEV would move away from the maximal value, i.e. $(0,x)$ with $0<x<1/2$, the unbroken gauge group would be the SM one, hence leading to a potentially interesting model. Hence, a realistic aGUT could be constructed if the potential can be destabilised via the addition of more fields and/or interactions, leading to a non-minimal model.

\subsection{An $E_7$ example}
\label{E7example}

Having three independent parities $E_7$ offers more possibilities in terms of symmetry-breaking patterns. 
Among the possibilities listed in Table~\ref{tab:Olist}, two are potentially interesting for aGUT model building.

The first corresponds to the $\SU(4)\times\SU(4)\times\U(1)$ obtainable via $\mathcal{P}_1\times\mathcal{P}_2$. While QCD colour has been extended to $\SU(4)$ in the Pati-Salam fourth-colour model \cite{Pati:1974yy}, attempts to extend the $\SU(2)_L$ to $\SU(4)$ have also been explored in the literature \cite{Foot:1994ym,Pisano:1994tf,Sanchez:2008qv,Riazuddin:2008yx}, showing some interesting features for leptons. However, in the $E_7$ embedding, the fundamental only contains bi-fundamentals. This makes it impossible to embed the SM gauge symmetry within the $\SU(4)\times\SU(4)$ structure while obtaining the correct SM fermions \cite{Bernardini:2007ui,Cembranos:2019yio}. Hence, this case can be discarded. For completeness, we checked that the gauge-scalar potential prefers the $\SU(4)\times\SU(4)\times\U(1)$ alignment; see Appendix~\ref{appE7}.

The second possibility is offered by the $\SU(6)\times\SU(2)\times\U(1)$ pattern, which can be obtained by all three non-trivial parity combinations. The idea would be to embed $\SU(4)\times\SU(2)_L$ of Pati-Salam within $\SU(6)$, while the remaining $\SU(2)$ is identified with $\SU(2)_R$. This setup would lead, technically, to a non-minimal aGUT as the quantitative unification of two SM couplings is required. Nevertheless, we discuss this case to exhaust all possible exceptional model-building cases, and because it offers some interesting features. The SM fermions must be embedded in the following two representations of $\SU(6)\times\SU(2)_R$:
\begin{eqnarray}
    (15,1) &\to & (6,1,1) \oplus (4,2,1) \oplus (1,1,1)\,, \\
    (6,2) &\to & (4,1,2) \oplus (1,2,2)\,,
\end{eqnarray}
where we show the decomposition under the PS group. The correct zero mode structure, however, can only be obtained in the model based on $\mathcal{P}_2\times\mathcal{P}_3$, for which a second possible alignment is present, see Table~\ref{tab:Olist}. Hence, we first study the stability of this orbifold.

\begin{figure}[H]
    \centering
    \includegraphics[height=6.3cm]{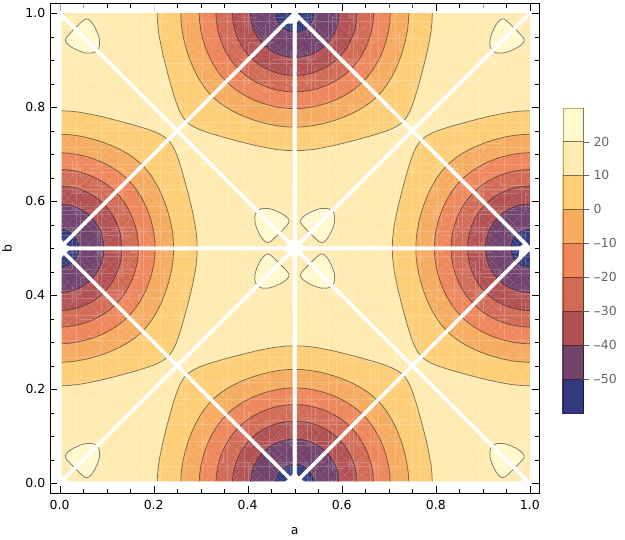} \hfill \includegraphics[height=6.3cm]{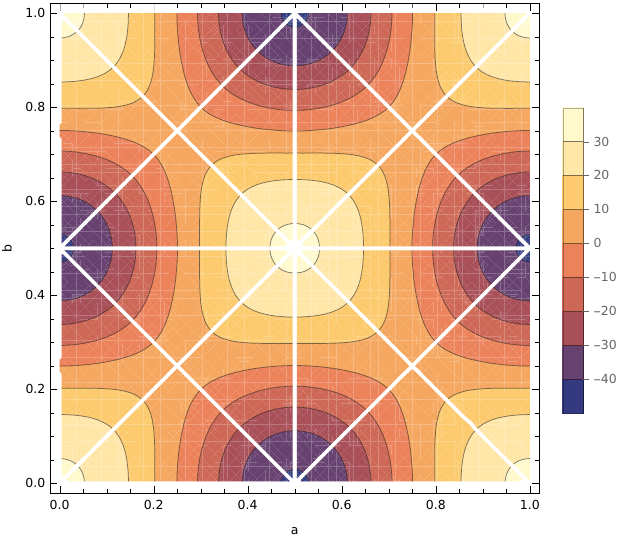}
    \caption{Plot of the gauge contributions to the effective potential (left) and the full effective potential (right) as a function of the two VEVs $a$ and $b$ for the $E_7$ model.}
    \label{fig:PotentialE72VEVS}
\end{figure}

The two parities break $E_7$ as follows (c.f. Table~\ref{tab:Plist}):
\begin{eqnarray}
    \mathcal{P}_2 &\Rightarrow & E_7 \to \SO(12) \times \SU(2)_R\,, \\
    \mathcal{P}_3 &\Rightarrow & E_7 \to E_6 \times \U(1)_X\,.
\end{eqnarray}
The two alignments listed in Table~\ref{tab:Olist} lead to the following unbroken groups:
\begin{itemize}
    \item[-] $\SO(10) \times \U(1)_\psi \times \U(1)_X$, where $\SO(10) \times \U(1)_\psi \subset E_6$. The fundamental and adjoint of $E_7$ decompose as follows:
    \begin{eqnarray}
        {\bf 56} &\to & 16^{+-}_{1,-2} \oplus 10^{++}_{-2,-2} \oplus 1^{++}_{4,-2} \oplus \overline{16}^{--}_{-1,2} \oplus 10^{-+}_{2,2} \oplus 1^{-+}_{-4,2} { \oplus 1_{0,6}^{+-} \oplus 1_{0,-6}^{++} } \,, \\
        {\bf 133} &\to & 45^{++}_{0,0} \oplus 1^{++}_{0,0} \oplus 1^{++}_{0,0} \oplus 16^{+-}_{-3,0} \oplus \overline{16}^{+-}_{3,0} \oplus 16^{--}_{1,4} \oplus \overline{16}^{--}_{-1,-4} \oplus \nonumber \\
        &&10^{-+}_{-2,4} \oplus 1^{-+}_{4,4} \oplus 10^{-+}_{2,-4} \oplus 1^{-+}_{-4,-4}\,.
    \end{eqnarray}
    Hence, the gauge-scalar transforms as $\varphi_{\SO(10)} = 16_{1,4} + \mbox{c.c.}$.

    \item[-] $\SU(6)\times\SU(2)_R\times\U(1)_X$ where $\SU(6)\times\SU(2)_R\subset E_6$ and $\SU(6)\times\U(1)_X \subset \SO(12)$. The fundamental and adjoint of $E_7$ decompose as:
    \begin{eqnarray}
        {\bf 56} &\to & (15,1)^{-+}_{-2} \oplus (1,1)^{--}_{6} \oplus (6,2)^{+-}_{2} \oplus (\overline{15},1)^{--}_{2} \oplus (1,1)^{-+}_{-6} \oplus (\overline{6},2)^{++}_{-2}\,, \\
        {\bf 133} &\to & (35,1)^{++}_{0} \oplus (1,3)^{++}_{0} \oplus (1,1)^{++}_{0} \oplus (15,1)_{4}^{+-} \oplus (\overline{15},1)^{+-}_{-4} \oplus \nonumber \\
        && (6,2)^{--}_{-4} \oplus (\overline{6},2)^{--}_{4} \oplus (20,2)^{-+}_{0}\,.
    \end{eqnarray}
    Hence, the gauge-scalar transforms as $\varphi_{\SU(6)} = (6,2)_{-4} + \mbox{c.c.}$. 
\end{itemize}

We computed the gauge-scalar potential for the latter, see Appendix~\ref{appE7} for more details. Two vacuum expectation values are allowed, which we label $a$ and $b$, leading to the following potential contribution from the gauge multiplet:
\begin{equation}
    \mathcal{V}_{\rm Adj} = \frac{3}{2} \left[F^+ (2a) + F^+ (2b) \right] + 2 \left[ F^+ (a+b) + F^+ (a-b)\right] + 6 \left[ F^- (a+b) + F^- (a-b)\right]\,.
\end{equation}
This potential is shown in the left panel of Fig.~\ref{fig:PotentialE72VEVS}, and it features global minima at $(0,1/2)$ and $(1/2,0)$. These configurations break $\SU(6)\times\SU(2)_R \to \SU(5) \times \U(1)$, however, additional zero modes appear that reconstruct invariance under $\SO(10)$. Hence, the stable orbifold is the $\SO(10)\times\U(1)^2$ case, similar to what we observed for the $E_6$ case.

Before concluding the discussion on this model, we need to include the effect of bulk matter fields. A complete SM generation can be embedded into a single bulk fermion ${\bf 56}$ with parities $(+,-)$, so that the field $\Psi_{\bf 56}^{+-}$ contains a right-handed zero mode $(15,1)_{-2}$ and a left-handed one $(6,2)_{2}$, with an additional right-handed singlet $(1,1)_{-6}$. The gauge scalar decomposes as
\begin{equation}
    \varphi_{\SU(6)} \to (4,1,2)_{-4} \oplus (1,2,2)_{-4} + \mbox{c.c.}
\end{equation}
so that it contains a Higgs doublet candidate (the second term) together with a scalar that could break the PS symmetry to the SM one. Hence, the SM Yukawa couplings are generated directly from gauge interactions, as in Gauge-Higgs unification models. The only bulk coupling, the gauge one, has a fixed point
\begin{equation}
    \alpha_g^\ast = \frac{2\pi}{63-8 n_g}\,,
\end{equation}
for $n_g\leq 7$ fermion generations. To check the orbifold stability, we included the contribution of the bulk fermions:
\begin{equation}
    \mathcal{V}_{\rm ferm.} = \frac{3}{16} \left[ F^+ (2a+2b) + F^+ (2a-2b)\right]+\frac{3}{4} \left[ F^+ (2a) + F^+ (2b)\right]\,.
\end{equation}
The total potential for three families is illustrated in the right panel of Fig.~\ref{fig:PotentialE72VEVS}, confirming that the global minima prefer the alignment $\SO(10)\times\U(1)^2$. We also remark that destabilising the potential for the gauge-scalar would not lead to a feasible model, as it can only break $\SU(6)\times\SU(2)_R$ to a flipped $\SU(5)$ model.

Finally, we also note that this $E_7$ case contains the $E_6$ orbifold given by $\mathcal{P}_0\times\mathcal{P}_2$, c.f. Table~\ref{tab:Olist}. The main difference is that the $E_6$ case is automatically stable on the breaking of $\SU(6)\times\SU(2)$ without any gauge-scalar. Hence, the Higgs and the scalar responsible for the breaking of $\SU(6)\times\SU(2)$ to Pati-Salam must be added in the bulk, leading to a non-minimal aGUT.

\section{Conclusions and perspectives} \label{sec:concl}

In this work we investigated the stability of orbifolds based on exceptional groups in 5D, compactified on $S^1/\mathbb{Z}_2\times\mathbb{Z}_2'$. We only studied alignments of the parities leading to maximal unbroken subgroups, as they always contain the stable configurations for $\SU(N)$, $\Sp(2N)$ and $\SO(N)$ cases \cite{Cacciapaglia:2024duu}. We applied such results to the construction of minimal aGUTs, following the prescriptions established in Refs.~\cite{Cacciapaglia:2023kyz,Cacciapaglia:2024duu}. Realistic models can only be built based on $E_6$ and $E_7$, as $E_8$ always leads to real unbroken subgroups in 5D while $G_2$ and $F_4$ do not contain the SM gauge symmetry. However, we demonstrated that all minimal aGUT models one could build, which fully embed the SM, are based on unstable orbifolds, hence they are not feasible.

Nevertheless, our exploration allows us to define possible non-minimal model-building avenues. The only promising one is based on $E_6$, broken by the only two un-equivalent parities, $\mathcal{P}_1\times\mathcal{P}_2$. The stable orbifold would break $E_6 \to \SU(5)\times\U(1)^2$, hence requiring quantitative unification before the extra dimension appears. However, we found that by turning on one vacuum expectation value for the gauge-scalar, this is connected to another alignment leading to $E_6 \to \SU(4)\times\SU(2)^2 \times\U(1)$: for non-extreme values, the unbroken symmetry is exactly that of the SM, hence a realistic model could be constructed if the one-loop potential for the gauge-scalar is suitably modified via non-minimal interactions.

Another possibility is still offered by the same $E_6$ orbifold, if the theory is supersymmetric. In such a case, the one-loop potential for the gauge scalar is vanishing, unless supersymmetry is broken. A feasible model has been constructed in Ref.~\cite{Cacciapaglia:2023ghp}, with the interesting feature that left-handed SM fermions arise from the gaugino fields and the Yukawa couplings are generated by gauge interactions. Hence, all couplings are guaranteed to flow towards the attractive UV fixed point. The stability of this orbifold is tightly related to the breaking of supersymmetry, and we leave a detailed study of this model for future publications. 

\section*{Acknowledgements}

ASC is supported in part by the National Research Foundation (NRF) of South Africa. Z.-W.W. is supported in part by the National Natural Science Foundation of China (Grant No.~12475105).

\appendix

\section{Computation of the effective potential}

We provide some details on the computation of the one-loop effective potential for the gauge-scalars in the non-trivial $F_4$, $E_6$ and $E_7$ orbifolds. The potential in Eq.~\eqref{eq:potgen} is fully determined by the dependence of the spectrum on the various VEVs of the gauge-scalar, i.e. on the interactions between the bulk fields and the fifth polarisation of the gauge field $A_5$. The most general procedure consists in defining the VEVs of  $A_5$ and then computing their effect on the various components of the bulk fields. To simplify the task, we employ a trick first proposed in Ref.~\cite{HABA2004166} in order to recast the computation in terms of known ordinary groups such as $\SU(N)$ and $\Sp(2N)$, for which general results are presented in Ref.~\cite{Cacciapaglia:2024duu}. The trick consists in identifying a subgroup $\mathcal{K}$ of the bulk group $\mathcal{G}$, satisfying the following two criteria:
\begin{itemize}
    \item[-] $\mathcal{K} \supset \mathcal{H}$, i.e. it contains the 4D unbroken group of the orbifold;
    \item[-] the parities defining the orbifold break $\mathcal{K} \to \mathcal{H}$ on both boundaries, so that the adjoint of $\mathcal{K}$ contains the zero mode gauge-scalars.
\end{itemize}
Hence, one can decompose each representation of $\mathcal{G}$ in representations of $\mathcal{K}$, and compute the contribution to the potential of each component. We will see how this works in the following examples.

\subsection{The $F_4$ orbifold}\label{app:F4}

As outlined in Table~\ref{tab:Olist}, the only non-trivial combination of parities breaks $F_4$ to $\SO(9)$ on one boundary and to $\Sp(6)_1\times \SU(2)_2$ on the other, leading to the 4D remnant group $\mathcal{H} = \Sp(4)\times \SU(2)_1\times\SU(2)_2$, where $\Sp(4)\times\SU(2)_1 \subset \Sp(6)_1$. Under $\mathcal{H}$, the adjoint of $F_4$ decomposes as: 
\begin{equation}
    {\bf{52}} \to (10,1,1)^{++} \oplus (1,3,1)^{++}\oplus (1,1,3)^{++} \oplus(5,2,2)^{+-} \oplus (4,1,2)^{--} \oplus (4,2,1)^{-+}\,,
    \label{eq:F4adjdecomp}
\end{equation}
with a gauge scalar in the $(4,1,2)$ representation.

We note that the components with parities $(\pm,\pm)$ can be accommodated in an adjoint of an $\Sp(6)$ group, leading to the identification of 
\begin{equation}
    \mathcal{K} = \Sp(6)_2\times\SU(2)_1\,,
\end{equation}
where $\Sp(6)_2\supset\Sp(4)\times\SU(2)_2$. Under $\mathcal{K}$, the gauge-scalar is part of the adjoint of $\Sp(6)_2$, and the decomposition of the $F_4$ adjoint will be given by:
\begin{equation}
    {\bf 52} \to  (1,3)^+ \oplus (21,1)^+ \oplus (14',2)^-\,.
    \label{eq:F4adjointdecomp}
\end{equation}
In the above notation, the first two terms in eq.~\eqref{eq:F4adjointdecomp} contain components with $(\pm,\pm)$ parities, while the last term includes the $(\pm,\mp)$ contributions.
From Eq.~\eqref{eq:F4adjointdecomp}, we can schematically write the gauge contribution to the effective potential:
\begin{eqnarray}
    \left.\mathcal{V}_{\rm Adj} \right|_{F_4}=\mathcal{V}_{21}+2\ \mathcal{V}_{14'}\,.
\end{eqnarray}
The two terms can now be computed using the results for the breaking of $\Sp(6) \to \Sp(4)\times \Sp(2)$. The gauge-scalar VEV can be parametrised in terms of a single non-zero entry, $a$, and following Ref.~\cite{Cacciapaglia:2024duu}, we obtain:
\begin{equation}
    \mathcal{V}_{21}(a)=2\ \left[{F}^+ (2a)+2{F}^{+} (a)\right]\,,
\end{equation}
while the $14'$ with negative parity contributes as:
\begin{equation}
    \mathcal{V}_{14'}(a)=2\ {F}^{-} (a)\,.
    \label{eq:14'potential}
\end{equation}
For Eq.~\eqref{eq:14'potential} we took into account the opposite parity of this representation with respect to the $\Sp(6)$ adjoint.
The total gauge contribution to the potential then becomes:
\begin{equation}
    \left.\mathcal{V}_{\rm Adj}\right|_{F_4}(a)=\frac{9}{4} \ F^+ (2a)\,.
\end{equation}
The global minimum sits at $a=0$, thus making this case stable.

\subsection{The $E_6$ orbifold} \label{app:E6}
The relevant combination of parities stems from $\mathcal{P}_1$ breaking $E_6 \to \SO(10) \times \U(1)$ on one boundary, and $\mathcal{P}_2$ breaking $E_6 \to \SU(6) \times \SU(2)_R$ on the other boundary, see Table~\ref{tab:Olist}. There are two possible maximal 4D remnant groups:  $\SU(5)\times \U(1)^2$ and $\SU(4)\times \SU(2)_L\times \SU(2)_R\times \U(1)$. 

For aGUT model-building purposes, we are interested in the latter case, under which the adjoint of $E_6$ decomposes according to Eq.~\eqref{eq:78deco}, with a gauge-scalar in the $(4,1,2)_3$+c.c. representation. We remark that $\SU(6)\supset\SU(4)\times\SU(2)_L$.
To compute the gauge-scalar potential, the group broken on both boundaries, which also contains this gauge-scalar, is given by:
     \begin{equation}
         \mathcal{K} = \SU(6)_R\times \SU(2)_{L}\,.
     \end{equation}
It is, therefore, enough to study the breaking of $\SU(6)_R\to \SU(4)\times\SU(2)_R$, for which general formulae can be found in Refs.~\cite{Naoyuki,Cacciapaglia:2024duu}. Under $\mathcal{K}$, the $E_6$ adjoint decomposes as:
     \begin{equation}
         {\bf 78} \to  (1,3)^{+} \oplus (35,1)^{+} \oplus (20,2)^-\,,
         \label{eq:78decomposition}
     \end{equation}
    Similarly to the $F_4$ case, the first two terms include the $(\pm,\pm)$ parity states, while the third contains $(\pm,\mp)$ contributions. The gauge contribution to the potential can then be written as:
     \begin{eqnarray}
         \left.\mathcal{V}_{\rm Adj}\right|_{E_6} =\mathcal{V}_{35}+2\ \mathcal{V}_{20}\,.
     \end{eqnarray}
     The gauge-scalar allows for two independent VEVs, $a$ and $b$.
     The contributions $\mathcal{V}_{35}$ and $\mathcal{V}_{20}$ to the effective potential coming from the adjoint $35$ and the 3-index antisymmetric $20$ representations of $\SU(6)$  are computed explicitly in Refs.~\cite{Cacciapaglia:2024duu, Naoyuki}:
     \begin{equation}
         \mathcal{V}_{35}(a,b)= F^+ (2a)+ F^+ (2b)+ 2\ \left[F^+ (a+b)+F^+ (a-b)\right] +4\ \left[ F^+ (a)+ F^+ (b)\right]\,,
         \label{eq:potential35}
     \end{equation}
     \begin{equation}
         \mathcal{V}_{20}(a,b)=2\ \left[ F^- (a+b)+ F^- (a-b)+ F^- (a)+ F^- (b)\right]\,.
         \label{eq:potential20}
     \end{equation}
     As before, we notice the presence of $F^-$ functions in Eq.~\eqref{eq:potential20}, compared to $F^+$ in Eq.~\eqref{eq:potential35}, due to the opposing parities of these representations in Eq.~\eqref{eq:78decomposition}.
     Adding them together, the total gauge contribution is found to be:
\begin{equation}
\begin{split}
    \left.\mathcal{V}_{\rm Adj}\right|_{E_6}^{\rm gauge}(a,b) = & \frac{5}{4} \left[ F^+(2a) +F^+(2b) \right] + \frac{1}{8} \left[ F^+(2a+2b) + F^+(2a-2b) \right]  \\
    & + 2 \left[ F^-(a+b)+F^-(a-b) \right]\,,
\end{split}
\end{equation}
with global minima at $(a,b)=(1/2,0)$ and $(0,1/2)$, hinting towards the instability of this orbifold.

As mentioned in Subsection~\ref{subsec:E6}, the fermionic and scalar degrees of freedom reside in separate fundamental $\bf{27}$ representations. To find the corresponding potential, we decompose the $\bf{27}$ under $\mathcal{K}$:
\begin{equation}
    {\bf 27}\rightarrow (\overline{6},2)^{+}\oplus(15,1)^{-}\,,
\end{equation}
such that the potential generated by this representation becomes:
\begin{equation}
    \mathcal{V}_{27}=2\ \mathcal{V}_{6}+\mathcal{V}_{15}\,.
\end{equation}
Using the $\SU(N)$ general formulae \cite{Cacciapaglia:2024duu, Naoyuki} for the fundamental $6$ and the 2-index antisymmetric $15$ of $\SU(6)$, we find the contribution of a field in the fundamental ${\bf 27}$ of $E_6$ with parities $(+,+)$ to be given by:
\begin{equation}
    \mathcal{V}_{ 27}(a,b)=F^-(a+b)+F^-(a-b)+2 \left[F^+(a)+F^+(b)+F^-(a)+F^-(b)\right]\,.
\end{equation}
The above formula is valid for bulk fields with overall parities $(+,+)$ and $(-,-)$, as it is the case for one bulk fermion and the bulk scalar. For bulk fields with overall parities $(+,-)$ and $(-,+)$, it suffices to exchange the functions $F^+ \leftrightarrow F^-$.
Hence, in the model, the bulk scalar contribution reads:
\begin{equation}
  \mathcal{V}_{\rm scalar} (a,b)  = 2 \left[\frac{1}{8} \left[ F^+(2a)+F^+(2b) \right] + \left[ F^-(a+b) + F^-(a-b) \right]\right]\,,   
\end{equation}
where the factor of $2$ accounts for the complexity of the field.
The fermionic content is embedded in two copies of the fundamental $\bf{27}$, with parities $(+,+)$ and $(-,+)$ per generation, giving an overall contribution:
\begin{eqnarray}
    \mathcal{V}_{\rm fermion}(a,b) = \frac{n_g}{4} \left[ F^+(2a)+F^+(2b)+\frac{1}{4} \left[ F^+(2a+2b) + F^+(2a-2b) \right] \right]\,.
\end{eqnarray}
Combining the contributions as in Eq.~\eqref{eq:potgen}, the total effective potential for the $E_6$ aGUT model still has global minima at $(a,b) = (1/2,0)$ and $(0,1/2)$.

To see what symmetry-breaking pattern occurs at the minimum of the potential, we need to study the effect of the gauge-scalar VEV on the zero-mode spectrum of the gauge fields. Firstly, it is clear that turning on a single VEV would break
\begin{equation}
\SU(4)\times\SU(2)_L\times\SU(2)_R\times\U(1)_\psi \to \SU(3)\times\SU(2)_L\times\U(1)^2\,,
\end{equation}
that is to the SM gauge group with an additional $\U(1)$ factor. Next, we focus on the components of the gauge multiplet that have parities $(\pm,\mp)$: they have masses given by $n+1/2$ in units of the extra dimension radius; hence a shift by $1/2$ given by the maximal VEV would generate zero modes. From the gauge coupling, we see that there is only one relevant coupling of the gauge-scalar, written as follows:
\begin{equation}
    A_5 A_\mu A^\mu \supset (\overline{4},1,2)_{-3}^{--}\ (6,2,2)_0^{+-}\ (\overline{4},2,1)_3^{-+} + \mbox{h.c.} 
\end{equation}
where the components stem from the corresponding field, i.e. the first stems from $A_5$ while the other two from the vectors. The gauge-scalar VEV, therefore, couples with a $(3,2)$ state under the SM gauge group and induces a new zero mode in such a state once the VEV is maximally equal to $1/2$. This new zero-mode gauge boson, therefore, reconstructs the gauge group $\SU(5)$, hence indicating that the gauge symmetry at the minimum of the potential is enlarged to $\SU(5)\times\U(1)^2$.

\subsection{The $E_7$ orbifolds} \label{appE7}
\subsubsection{$\mathcal{P}_{2}\times \mathcal{P}_3$ breaking }
The parity combination in this case consists of  $\mathcal{P}_2$, breaking $E_7 \to \SO(12) \times \SU(2)_R$ on one boundary, and $\mathcal{P}_3$ , breaking $E_7 \to E_6 \times \U(1)_X$ on the other. The maximal subgroups, c.f. Table~\ref{tab:Olist}, are $\SU(6)\times \SU(2)_R\times \U(1)_X$ and $\SO(10)\times \U(1)_\psi\times \U(1)_X$. 

For the potential computation, it is most convenient to use the former case, for which the gauge-scalar is in the $(6,2)_{-4}$ representation of the unbroken group $\mathcal{H}$. The group broken on both boundaries is 
\begin{equation}
    \mathcal{K} = \SU(8)\,,
\end{equation}
which contains the gauge-scalar in its adjoint representation. We can decompose the $E_7$ adjoint in this basis to find:
\begin{equation}
    {\bf 133} \to 63^+ \oplus 70^-\,,
\end{equation}
where the $\SU(8)$ adjoint contains the $(\pm,\pm)$ states, while 4-index antisymmetric the $(\pm,\mp)$ ones. We write the gauge contribution to the effective potential in a similar fashion:
\begin{equation}
    \left.\mathcal{V}_{\rm Adj}\right|_{E_7} =  \mathcal{V}_{63} + \mathcal{V}_{70}\,,
\end{equation}
where $\mathcal{V}_{63}$ is the contributions from the $\SU(8)$ adjoint and $\mathcal{V}_{70}$ from the $4$-index antisymmetric representation $70$. The first term can be computed with the general formula for two VEVs, $a$ and $b$ \cite{Cacciapaglia:2024duu,Naoyuki}:
 \begin{equation}
      \mathcal{V}_{63}(a,b)= F^+(2a) + F^+(2b) + 2\ \left[ F^+(a+b) + F^+(a-b)\right] + 8\ \left[ F^+(a) + 8 F^+(b)\right]\,.
  \end{equation}
  The contribution from the $4$-index antisymmetric representation $70$ yields:
  \begin{equation}
      \mathcal{V}_{70}(a,b)=6\ \left[ F^-(a+b) +  F^-(a-b)\right]+8\ \left[F^-(a) + F^-(b)\right]\,.
  \end{equation}
  We thus find the gauge contribution to the effective potential to be:
  \begin{equation}
  \begin{split}
       \left.\mathcal{V}_{\rm Adj}\right|_{E_7} (a,b)  = &  \frac{3}{2} \left[F^+(2a)+F^+(2b) \right] + 2\ \left[ F^+(a+b)+F^+(a-b) \right] \\
       &  +6\ \left[F^-(a+b)+F^-(a-b) \right]\,.
  \end{split}
  \end{equation}
This potential has minima at $(a,b) = (1/2,0)$ and $(0,1/2)$.

As mentioned in Section \ref{E7example}, a potential non-minimal aGUT would contain a family of SM fermions within a single fundamental representation of $E_7$ with parities $(+,-)$, $\Psi_{\bf 56}^{+-}$. No bulk scalar is needed, as the SM Higgs field emerges from Gauge-Higgs unification. The contributions to the effective potential coming from $\Psi_{\bf 56}^{+-}$ can be computed by considering the decomposition of ${\bf 56}$ in terms of the $\SU(8)$ basis:
\begin{equation}
    {\bf 56} \to  28^+ \oplus \overline{28}^-\,,
\end{equation}
    where $28$ is the $2$-index antisymmetric representation of $\SU(8)$, and the two components have opposite parities. The contribution of a single $28$ gives:
    \begin{equation}
        \mathcal{V}_{28}(a,b)=4\left[F^+(a)+F^+(b)\right]+F^+(a+b)+F^+(a-b)\,,
    \end{equation}
  such that, taking into account the opposite parities,
  \begin{equation}
      \mathcal{V}_{\bf 56}(a,b)=\frac{1}{4}\left[F^+(2a)+F^+(2b)\right] + \frac{1}{16}\left[F^+(2a+2b)+F^+(2a-2b)\right]\,.
  \end{equation}
Adding this contribution to the potential leaves the global minima at the same places, $(1/2,0)$ and $(0,1/2)$.

As for the $E_6$ orbifold, one can check that at the minimum, $\SU(6)\times\SU(2)_R \to \SU(5)\times \U(1)^2$, while additional zero modes complete an unbroken group $\SO(10) \supset \SU(5)\times\U(1)$, so that the stable orbifold corresponds to the unbroken maximal group $\SO(10)\times\U(1)^2$.

\subsubsection{$\mathcal{P}_{1}\times \mathcal{P}_2$ breaking }

In this case, the orbifold is defined by the parities  $\mathcal{P}_1$, breaking $E_7 \to \SU(8)$ on one boundary and $\mathcal{P}_2$ , breaking $E_7 \to \SO(12) \times \SU(2)$ on the other. The maximal groups are $\SU(4)\times \SU(4)\times \U(1)_X$ and $\SU(6)\times \SU(2)\times \U(1)_Z$, where $\SU(4)\times\SU(4)\subset\SO(12)$ and $\SU(6)\times\U(1)_Z \subset \SO(12)$. 

For the potential computation, it is most convenient to consider the former case, with $\mathcal{H} = \SU(4)\times\SU(4)\times\U(1)_X$.
The decomposition of the adjoint under $\mathcal{H}$ is:
        \begin{equation}
\begin{split}
    {\bf 133} \to & (15,1)_0^{++}\oplus (1,15)_0^{++}\oplus (1,1)_0^{++}\oplus (4,\bar{4})_2^{+-} \oplus  (\bar{4},4)_{-2}^{+-} \oplus (6,6)_0^{-+} \\
    &  \oplus (1,1)_{-4}^{-+}+(1,1)_4^{-+} \oplus (4,\bar{4})_{-2}^{--} \oplus (\bar{4},4)_{2}^{--}\,,
\end{split}
\end{equation}
with gauge-scalars in the $(4,\bar{4})_{-2}$ and $(\bar{4},4)_{2}$ representations.
The group broken on the two boundaries, and whose adjoint contains the gauge-scalar, is 
\begin{equation}
    \mathcal{K} = \SU(8)'\,,
\end{equation}
misaligned to the $\SU(8)$ preserved by $\mathcal{P}_1$.
Under $\mathcal{K}$, the adjoint of $E_7$ decomposes as:
\begin{equation}
    {\bf 133} \to 63^+ \oplus 70^-\,,
\end{equation}
where the $63$ and $70$ dimensional representations are the adjoint and the $4$-index antisymmetric of $\SU(8)$, respectively. As for the previous case, the gauge contribution to the effective potential can be written:
\begin{equation}
    \left.\mathcal{V}_{\rm Adj}\right|_{ E_7} = \mathcal{V}_{63} + \mathcal{V}_{70}\,. 
\end{equation}
We can, once again, employ the general $\SU(N)$ formulae \cite{Cacciapaglia:2024duu,Naoyuki} to compute the potential, which will now be a function of four VEVs, $a_i$ with $i=1,2,3,4$. This leads to the following partial contributions:
\begin{equation}
    \mathcal{V}_{63}(a_i) = \sum_{i=1}^4 F^+(2a_i) +2\ \sum_{ij} \left[F^+(a_i+a_j)+ F^+(a_i-a_j) \right]\,,
\end{equation}
and
\begin{equation}
\begin{split}
     \mathcal{V}_{70}(a_i) = & F^+(a_1+a_2+a_3+a_4) + \sum_{ijlk} F^+(a_i+a_j+a_k-a_l) + \sum_{ijlk} F^+(a_i+a_j-a_l-a_k) \\ & + 2 \sum_{ij} \left[F^+(a_i+a_j) + F^+(a_i-a_j)\right]\,,
\end{split}
\end{equation}
where $\sum_{ij}$ and $\sum_{ijlk}$ contain all uniquely recurring combinations of the indices. Taking into account the opposite parity of the $70$ with respect to the $63$, the total gauge potential reads:
\begin{equation}
    \begin{split}
        \left.\mathcal{V}_{\rm Adj}\right|_{E_7} & (a_i) = 
         \sum_i F^+(2a_i)+\frac{1}{8} \sum_{ij} \left[F^+(2a_i+2a_j)+ F^+(2a_i-2a_j)\right] + \\ & F^-(a_1+a_2+a_3+a_4) + \sum_{ijlk} F^-(a_i+a_j+a_l-a_k) + \sum_{ijlk} F^-(a_i+a_j-a_l-a_k)\,.
    \end{split}
\end{equation}
This potential has a global minimum at $a_i=0$, proving that this is the stable orbifold symmetry breaking pattern. 

\bibliographystyle{utphys}
\bibliography{biblio}

\end{document}